\documentclass[a4paper,11pt]{article}

\pdfoutput=1 

\usepackage{jcappub} 
\usepackage{amsmath}
\usepackage{amsfonts}
\usepackage{amssymb}
\usepackage{graphicx}

\makeatletter
\def\@hangfrom#1{\setbox\@tempboxa\hbox{{#1}}%
      \hangindent 0pt
      \noindent\box\@tempboxa}
\makeatother
\def\un#1{\relax\ifmmode\@@underline#1\else
        $\@@underline{\hbox{#1}}$\relax\fi}
\let\du=\du                     
\def\a{\alpha}

\def\d{\delta}

\def\f{\phi}

\def\m{\mu}
\def\n{\nu}

\def\s{\sigma}

\def\bo{{\raise-.3ex\hbox{\large$\Box$}}}               
\def\TH{{\raise.2ex\hbox{$\displaystyle \bigodot$}\mskip-4.7mu \llap H \;}}
\def\face{{\raise.2ex\hbox{$\displaystyle \bigodot$}\mskip-2.2mu \llap {$\ddot
        \smile$}}}                                      

\def\abs#1{\left| #1\right|}                    
\def\leftrightarrowfill{$\mathsurround=0pt \mathord\leftarrow \mkern-6mu
        \cleaders\hbox{$\mkern-2mu \mathord- \mkern-2mu$}\hfill
        \mkern-6mu \mathord\rightarrow$}
\def\dvec#1{\vbox{\ialign{##\crcr
        \leftrightarrowfill\crcr\noalign{\kern-1pt\nointerlineskip}
        $\hfil\displaystyle{#1}\hfil$\crcr}}}           
\def\frac#1#2{{\textstyle{#1\over\vphantom2\smash{\raise.20ex
        \hbox{$\scriptstyle{#2}$}}}}}                   
\def\sfrac#1#2{{\vphantom1\smash{\lower.5ex\hbox{\small$#1$}}\over
        \vphantom1\smash{\raise.4ex\hbox{\small$#2$}}}} 
\def\bfrac#1#2{{\vphantom1\smash{\lower.5ex\hbox{$#1$}}\over
        \vphantom1\smash{\raise.3ex\hbox{$#2$}}}}       
\def\afrac#1#2{{\vphantom1\smash{\lower.5ex\hbox{$#1$}}\over#2}}    
\def\[{\lfloor{\hskip 0.35pt}\!\!\!\lceil}
\def\]{\rfloor{\hskip 0.35pt}\!\!\!\rceil}

\def\du#1#2{_{#1}{}^{#2}}

\def\fracm#1#2{\hbox{\large{${\fracmm{{#1}}{{#2}}}$}}}

\def\un{\underline}
\def\fracmm#1#2{{{#1}\over{#2}}}

\def\low#1{{\raise -3pt\hbox{${\hskip 0.75pt}\!_{#1}$}}}

\newskip\humongous \humongous=0pt plus 1000pt minus 1000pt

\newif\ifdtup

\def\({\left(}
\def\){\right)}

\def\beq{\begin{equation}}
\def\eeq{\end{equation}}
\def\bea{\begin{eqnarray}}
\def\eea{\end{eqnarray}}
\newcommand{\be}{\begin{equation}}
\newcommand{\ee}{\end{equation}}
\newcommand{\nbe}{\begin{equation*}}
\newcommand{\nee}{\end{equation*}}

\newcommand{\lb}{\label}

\title{\boldmath Inflation in $F(R)$ gravity models revisited after ACT}

\author[a,b,1]{Sergei V.~Ketov\note{The corresponding author.},}
\author[c]{Ekaterina O.~Pozdeeva,}
\author[c]{Sergey~Yu.~Vernov}

\affiliation[a]{Department of Physics, Tokyo Metropolitan University,\\
1-1 Minami-ohsawa, Hachioji-shi, Tokyo 192-0397, Japan}

\affiliation[b]{Kavli Institute for the Physics and Mathematics of the Universe (WPI),
\\ The University of Tokyo Institutes for Advanced Study, Kashiwa 277-8583, Japan}

\affiliation[c]{Skobeltsyn Institute of Nuclear Physics, Lomonosov Moscow State University,\\
Leninskiye Gory~1, Moscow 119991, Russia}

\emailAdd{ketov@tmu.ac.jp}
\emailAdd{pozdeeva@www-hep.sinp.msu.ru}
\emailAdd{svernov@theory.sinp.msu.ru}

\keywords{inflation, modified gravity, slow-roll approximation}

\arxivnumber{2508.08927}

\abstract{The $F(R)$ gravity models of inflation are revisited in light of the recent observations of cosmic microwave background radiation by Atacama Cosmology Telescope (ACT) and DESI Collaboration. A detailed study of the evolution equations in the Jordan frame is given and a new description of the slow-roll approximation in the $F(R)$-gravity-based models of inflation  is proposed. It is found that all those models of inflation are significantly constrained by demanding a higher (than the Planck Telescope value) cosmological tilt $n_s$ of scalar perturbations and a positive running index $\alpha_s$ favored by ACT. It is not difficult to meet the ACT constraints on the scalar tilt $n_s$ by modifying the existing models of inflation, but simultaneously demanding a positive running $\alpha_s$ would rule out many of them. Using the proposed slow-roll approximation in the Jordan frame, we provide a new modification of the Starobinsky inflation model in the framework of $F(R)$ gravity, which satisfies all ACT constraints. An extension of our ACT-consistent  inflation model to the unified $F(R)$-gravity description of Starobinsky-like inflation and production of primordial black holes on a smaller scale is also proposed.}

\begin{document}
\maketitle
\flushbottom

\section{Introduction}

Cosmological inflation is a paradigm that proposes the existence of a quasi-de-Sitter phase in the early Universe before the radiation-dominated era (approximately between $10^{-36}$ s to $10^{-32}$ s). Inflation is characterised by an accelerating expansion of the Universe with a "Graceful Exit", and resolves major internal problems in the standard (Einstein-Friedmann) cosmology, such as the flatness problem, the horizon problem, the initial conditions problem, etc. The driver of inflation is supposed to be a scalar field called inflaton whose microscopic quantum fluctuations
become macroscopic (or classical) during inflation and seed the large-scale structure formation in the Universe. A success of inflation is not limited to the above because it also predicts a nearly scale-invariant and Gaussian spectrum of primordial
perturbations in very good agreement with precision observations of the cosmic microwave background (CMB) radiation \cite{Planck:2018jri,BICEP:2021xfz}. Inflation is still a small window into the very early Universe and very high energy physics, while the detailed mechanism of inflation is unknown despite many (over 300) theoretical models proposed in the literature~\cite{Martin:2024qnn}.

Single-field models of inflation can provide a simple description of viable inflation in terms of canonical inflaton field $\phi$ minimally coupled to gravity and having a potential $V(\phi)$. The variety of single-field inflation models is thus parameterised by their potentials, whereas CMB observations still allow infinitely many possible potentials, even though the latter have the vanishing measure in the space of all potentials. Therefore, data-driven selection of inflaton potentials needs to be complemented by fundamental theoretical principles and constraints to discriminate  between the CMB-compatible potentials.

A scalar field theory minimally coupled to Einstein gravity (called the Einstein frame) is known to be classically equivalent to modified $F(R)$ gravity (called the Jordan frame) whose Lagrangian is given by a smooth function $F(R)$ of the spacetime Ricci curvature scalar $R$, see e.g. Ref.~\cite{Maeda:1988ab}. Both descriptions are related by a Weyl transformation of spacetime metric. The relation between those frames is non-trivial, and it can be used for imposing restrictions on the inflaton potential from the gravity side. Moreover, a description of inflation in pure $F(R)$ gravity without extra matter uses only gravitational interactions with general covariance at work, while it can be technically more advantageous by using only analytical methods instead of numerical calculations. The famous example is the Starobinsky inflation model defined in the Jordan frame by the action~\cite{Starobinsky:1980te,Ketov:2025nkr}
\begin{equation}
\label{starS}
    S_{\rm Star.} = \fracmm{M^2_{\rm Pl}}{2}\int d^4 x \sqrt{-g } \left( R + \fracmm{1}{6m^2}R^2\right)~,
\end{equation}
where $m\sim 10^{-5}M_{\rm Pl}$ is the inflaton mass and $M_{\rm Pl}$ is the reduced Planck mass.
 In the Einstein frame, it leads to the following potential of canonical inflaton $\phi$:
\begin{equation}
\label{starV}
    V_{\rm Star.}(\phi) = \fracmm{3}{4}  M^2_{\rm Pl}m^2 \left[ 1 - \exp\left(-\sqrt{\fracm{2}{3}}\phi/M_{\rm Pl}\right)\right]^2
\end{equation}
that nicely fits the CMB observations by Planck/BICEP~\cite{Planck:2018jri,BICEP:2021xfz}, though the origin of the potential (\ref{starV}) without the reference to the action (\ref{starS}) becomes obscure. The Starobinsky model in the Jordan frame is not only geometrical by relying on General Relativity but also explains  the origin of the approximate flatness of the potential $V(\phi)$ during inflation due to the scale invariance of the $R^2$-action~\cite{Mishra:2019ymr}, and has
an attractor solution for inflation too, which sharply distinguishes the potential (\ref{starV}) among others.

At the same time, the Starobinsky description of inflation is expected to provide merely the leading (or dominant) terms in the gravitational effective action and in the inflaton potential, respectively, because both of them are subject to corrections from the higher-order terms that are more difficult to calculate. On the one hand, those corrections are important to account for more precise CMB measurements in the future. On the other hand, a more experimental precision requires a more precise theoretical framework for the CMB description. As a result of all that, single-field models of inflation become more and more constrained. The $F(R)$ gravity models of inflation, generalizing the Starobinsky model, were well studied in the literature, see e.g., Refs.~\cite{Berkin:1990nu,Saidov:2010wx,Kaneda:2010ut,Kaneda:2010qv,Ketov:2010qz,Huang:2013hsb,Kaneda:2015jma,Miranda:2017juz,Ivanov:2021chn,Pozdeeva:2022lcj,Addazi:2025qra,Gialamas:2025ofz,Fazzari:2025nfr}. More recently, interest to those inflation models renewed due to the new data from the Atacama Cosmology Telescope (ACT)~\cite{ACT:2025fju,ACT:2025tim} and DESI 2024~\cite{DESI:2024mwx}.

The single-field inflation models become even more constrained when one adds a production of primordial black holes (PBH) at smaller scales \cite{Novikov:1967tw,Zeldovich:1967lct,Hawking:1971ei}. The standard mechanism of PBH production is via  engineering a nearly-inflection point of the potential $V$ \cite{Ivanov:1994pa,Garcia-Bellido:1996mdl,Germani:2017bcs,Ezquiaga:2017fvi,Chataignier:2023ago}. However, less is known about PBH production with inflation in the Jordan frame, which may provide additional insights. Quantum loop corrections might invalidate many single-field models of inflation with PBH production \cite{Kristiano:2022maq}, while loop calculations rely on a detailed shape of the potential $V$ \cite{Franciolini:2023agm} or that of the $F$-function \cite{Saburov:2023buy,Saburov:2024und}.

In this paper, we propose a new inflation model in $F(R)$ gravity that satisfies all the ACT/DESI data. To construct the model, we develop a new version of the slow-roll approximation in the Jordan frame.

Our paper is organised as follows. In section~2, a detailed study of the evolution equations in a generic $F(R)$ gravity theory is given. The correspondence between the Jordan and Einstein frames is systematically described in section~3. The inflation parameters are introduced and computed in sections~4 and 5. Also in section~5, we propose new explicit formulae for the slow-roll approximation in the Jordan frame. The Starobinsky model  of inflation and its predictions for the CMB data are reproduced in the Jordan frame in section~6 from the viewpoint of the proposed SR approximation. The CMB measurements by Planck/BICEP Collaborations~\cite{BICEP:2021xfz} and Atacama Cosmology Telescope (ACT) \cite{ACT:2025fju,ACT:2025tim} combined with  DESI 2024 data \cite{DESI:2024mwx} are discussed in section~7. A new $F(R)$ gravity model of Starobinsky inflation, extending the modified gravity Lagrangian in Eq.~(\ref{starS}) to a higher-order polynomial with respect to the spacetime scalar (Ricci) curvature and satisfying the ACT/DESI constraints within $1\sigma$, is presented in section~8. We extend our model to the unified description of inflation,  PBHs production and dark energy (DE) in section~9. We summarise our conclusions in section~10.

\section{Evolution equations in $F(R)$ gravity models}

Let us consider a generic $F(R)$ gravity model in a curved four-dimensional spacetime  with the action (in the Jordan frame)
\begin{equation}
\label{actionFR}
    S_F=\int d^4 x \sqrt{-g }F(R ),
\end{equation}
where $F$ is a (multi)differentiable function  of the  Ricci scalar $R$.

A variation of  the action (\ref{actionFR}) with respect to the metric $g_{\mu\nu}$ gives the following equations of motion:
\begin{equation}
F_{,R }R^\mu_{\nu}-\fracmm{1}{2}F\delta^\mu_\nu- \left(g^{\mu\rho} D_\rho\partial_\nu-\delta^\mu_{\nu}
\Box\right)F_{,R }=0\,,
\label{equ_F_R}
\end{equation}
where $F_{,R }=\fracmm{dF}{dR }$ and the operator $\Box$ is the covariant d'Alembertian,
\begin{equation*}
\Box F_{,R }={}\fracmm{1}{ \sqrt{-g }}\partial_\nu \left(\sqrt{-g }g^{\mu\nu}\partial_\mu  F_{,R }\right).
\end{equation*}

Equation (\ref{equ_F_R}) leads to the trace equation
\begin{equation}
\label{FRtrace}
F_{,R }R -2F+3\Box F_{,R }=0\,.
\end{equation}

To avoid graviton as a ghost and scalaron (inflaton) as a tachyon, one has to impose the following restrictions on the function $F$~\cite{Starobinsky:2007hu}:
\be \lb{restr2}
F_{,R } >0 \quad {\rm and} \quad F_{,R R } >0\,.
\ee

In the spatially flat Friedmann-Lemaitre-Robertson-Walker (FLRW) universe with the metric
\begin{equation}\label{metric}
ds^2={}-dt^2+a^2(t)\left(dx^2+dy^2+dz^2\right)\,,
\end{equation}
 the Ricci scalar is given by
 \begin{equation}\label{RH}
    R =6\left(\dot{H} +2H^2\right),
 \end{equation}
 where $H =\dot{a}/a$ is the Hubble parameter and dots denote the time derivatives.

With the FLRW metric, the $(0,0)$-component of the equations of motion reads
\begin{equation}\label{equ00}
  3\left(\dot{H} +H ^2\right)F_{,R }-\fracmm{1}{2}F =3HF_{,R R }\dot{R }\,,
\end{equation}
while the trace equation (\ref{FRtrace}) is a consequence of Eq.~(\ref{equ00}).

To describe cosmic evolution during inflation, we use the e-folds number $N=-\ln(a/a_{e})$, where $a_{e}$ is a constant. Using the relation $\fracmm{d}{dt}=-H\fracmm{d}{dN}$,  we find that Eq.~(\ref{equ00}) is equivalent to
\begin{equation}\label{SYSTEMFRN}
\left(H^2\right)'=4H^2-\fracmm{R}{3}\,,\qquad {R\,}'={} - \fracmm{\left(R-6H^2\right)F_{,R}-F}{6H^2F_{,RR}},
\end{equation}
where the primes denote the derivatives with respect to $N$.

It is possible to rewrite Eq.~(\ref{equ00}) to the form similar to the $(0,0)$-equation in the Einstein frame, by using the procedure proposed in Ref.~\cite{Skugoreva:2014gka} and  the following functions:
\begin{equation}
\label{YR}
Y=\fracmm{M_{\mathrm{Pl}}}{\sqrt{2F_{,R }}}\left(H +\fracmm{F_{,R R }\dot{R} }{2F_{,R }}\right),
\end{equation}
\begin{equation}
\label{AVeff}
A=\fracmm{3M_{\mathrm{Pl}}^4 F_{,RR}^2}{4F_{,R }^3}\,,\qquad V_{\mathrm{eff}}=\fracmm{M_{\mathrm{Pl}}^4}{4F_{,R }^2}\left(R F_{,R }-F\right)\,.
\end{equation}
Equation~(\ref{equ00}) implies that these functions satisfy the equation
\begin{equation}
\label{equY}
3M_{\mathrm{Pl}}^2Y^2=\fracmm{A}{2}\dot{R}^2+V_{\mathrm{eff}}
\end{equation}
that can be verified as follows:
\begin{equation}
\label{equYdetail}
\begin{split}
3M_{\mathrm{Pl}}^2Y^2=&\fracmm{3M_{\mathrm{Pl}}^4}{2F_{,R }}\left(H^2+\fracmm{H F_{,R R }\dot{R}}{F_{,R}} +\fracmm{F_{,R R }^2}{4F^2_{,R }}\dot{R}^2\right)\\
=&\fracmm{3M_{\mathrm{Pl}}^4}{2F_{,R }}H^2+\fracmm{3M_{\mathrm{Pl}}^4}{2F_{,R }^2}\left[\left(\dot{H} +H ^2\right)F_{,R }-\fracmm{F}{6}\right]+\fracmm{A}{2}\dot{R}^2\\
=&\fracmm{A}{2}\dot{R} ^2+V_{\mathrm{eff}}\,.
\end{split}
\end{equation}

\section{Correspondence between Jordan and Einstein frames}

Let us demonstrate that the function $Y(t)$ is the Hubble parameter in the Einstein frame as a function of the cosmic time in the Jordan frame. The $F(R)$ gravity action (\ref{actionFR}) can be rewritten as
\begin{equation}
    S_{J}=\int d^4 x \sqrt{-g } \left[U(\s)R -V(\sigma)\right]\,,
\end{equation}
where the scalar field $\s$ has been introduced,
\begin{equation}
\label{UVsigma}
U(\s)=F_{,\s}(\s),\qquad V(\s)=\s F_{,\s}-F\,,
\end{equation}
and the commas denote the derivatives with respect to a given scalar field.
When $F_{,\s\s}(\s)\neq 0$, the field $\s$ can be eliminated via its algebraic equation of motion, $R=\s$, which yields back the action (\ref{actionFR}). It follows from the definitions of $U$ and $V$ that
\begin{equation}
\label{dequsigma}
V_{,\sigma}=\s U_{,\sigma}=\s F_{,\s\s}\,.
\end{equation}

The evolution equations derived from the action $S_{J}$ in the spatially flat FLRW universe are given by~\cite{Kamenshchik:2024kay}
 \begin{equation}
\label{equ00H}
6UH^2+6\dot U H=V
\end{equation}
and
\begin{equation}
\label{equ01}
2U\dot H-H\dot{U}+\ddot{U}=0.
\end{equation}

After the Weyl transformation of the metric from the Jordan frame to the Einstein frame~as
\be \lb{wtrans}
g^{(E)}_{\mu\nu}=\fracmm{2F_{,\s}(\sigma)}{M^2_{\mathrm{Pl}}}g _{\mu\nu}\,,
\ee
one gets the following action in the Einstein frame~\cite{Maeda:1988ab}:
\begin{equation}
\label{SE}
S_{E} =\int d^4x\sqrt{-g^{(E)}}\left[\fracmm{M^2_{\mathrm{Pl}}}{2}R_E-\fracmm{h}{2}{g^{(E)\mu\nu}}\partial_\mu{\sigma}\partial_\nu{\sigma}-V_{eff}\right],
\end{equation}
where
\begin{equation}
\label{hsigma}
h(\sigma)=\fracmm{3M^2_{\mathrm{Pl}}}{2F_{,\s}^2}F_{,\s\s}^2=\fracmm{2U(\s)}{M^2_{\mathrm{Pl}}}\,A(\s)
\end{equation}
and
\begin{equation}
\label{Ve}
V_{eff}(\s)=\fracmm{M_{\mathrm{Pl}}^4}{4F_{,\s }^2}\left(R F_{,\s }-F\right)=\fracmm{M_{\mathrm{Pl}}^4 V}{4U^2}~.
\end{equation}

The Hubble parameter in the Einstein frame $H_E$ as a function of the cosmic time in the Jordan frame $t$ can be expressed via $H$ and the function $U(\s)$ as follows~\cite{Pozdeeva:2025ied}:
\begin{equation}
\label{HEHJ}
H_E=\fracmm{d\ln(a_E)}{dt_E}=\fracmm{M_{\mathrm{Pl}}}{\sqrt{2U}}\left(H +\fracmm{\dot{U}}{2U}\right),
\end{equation}
where $t_E$ is the cosmic time in the Einstein frame,
\begin{equation}
\fracmm{dt_E}{dt}=\fracmm{\sqrt{2F_{,R }}}{M_{\mathrm{Pl}}}\,.
\end{equation}

Equations~(\ref{equ00H}) and (\ref{equ01}) can be rewritten in terms of $H_E$ as
\begin{equation}
\label{equHe}
3M_{\mathrm{Pl}}^2H_E^2=\fracmm{A}{2}{\dot{\s}}^2+V_{\mathrm{eff}},
\end{equation}
\begin{equation}
\label{Fr21Qm}
\dot H_E={}-\fracmm{A\sqrt{2U}}{2M_{\mathrm{Pl}}^3}\,{\dot{\s}}^2,
\end{equation}
where the functions $A(\s)$ and $V_{eff}(\s)$ have been defined in Eq.~(\ref{AVeff}).

Differentiating Eq.~(\ref{equHe}) with respect to the time $t$ and using Eq.~(\ref{Fr21Qm}), one yields
\begin{equation}
\label{KGequVeff}
\ddot\s={}-3\sqrt{\fracmm{2U}{M_{\mathrm{Pl}}^2}}H_E\dot{\s}-\fracmm{A_{,\s}}{2A}{\dot{\s}}^2-\fracmm{V_{\mathrm{eff},\s}}{A}.
\end{equation}

Using Eqs.~(\ref{UVsigma}) and (\ref{HEHJ}), one gets $H_E=Y$, so that Eq.~(\ref{equHe}) is equivalent to Eq.~(\ref{equY}). Equation~(\ref{Fr21Qm}) can now be rewritten to the form
\begin{equation}
\label{Fr21Y}
\dot Y={}-\fracmm{A\sqrt{2F_{,R }}}{2M_{\mathrm{Pl}}^3}\,\dot{R}^2\,.
\end{equation}
Therefore,  $Y(t)$ is a monotonically decreasing function.

Defining the canonical scalar field $\f$ instead of $\s$ by the field redefinition
\begin{equation}
\label{phidF}
\phi=\sqrt{\fracmm{3}{2}}M_{\mathrm{Pl}}\ln\left[\fracmm{2}{M^2_{\mathrm{Pl}}}F_{,\s}\right]
\end{equation}
yields the action $S_{E}$ in the standard (quintessence or scalar-tensor) form
\begin{equation}\label{ActionSe}
    S_E=\int  d^4x \sqrt{-g^E}\left[\fracmm{ M^2_{\mathrm{Pl}}}{2}R_E-\fracmm{1}{2}g^{E\m\n}\partial_\mu\phi\partial_\nu\phi-V_E(\phi)\right],
\end{equation}
where $V_E(\phi)=V_{eff}(\s(\phi))$.

A variation of the action (\ref{ActionSe}) gives the following equations of motion in the spatially flat FLRW universe:
\begin{equation}\label{FriedmannEq1}
3M^2_\mathrm{Pl}H_E^2=\fracmm{1}{2}\left(\fracmm{d\phi}{dt_E}\right)^2+V_E\,,
\end{equation}
\begin{equation}\label{FriedmannEq2}
\fracmm{dH_E}{dt_E}={}-\fracmm{1}{2M_\mathrm{Pl}^2}\left(\fracmm{d\phi}{dt_E}\right)^2\,,
\end{equation}
and
\begin{equation}\label{fieldEq}
\fracmm{d^2\phi}{dt_E^2}+3H_E\fracmm{d\phi}{dt_E}+V_{E,\phi}=0\,.
\end{equation}

\section{Inflation parameters}

In this Section, we introduce  the Hubble flow parameters without any approximation, and then relate them to the cosmological tilts in the SR approximation, both in the Einstein frame and in the Jordan frame.

\subsection{Einstein frame}

The Hubble flow parameters in the Einstein frame are defined by~\cite{Liddle:1994dx}
\begin{equation}\label{epsilon}
\varepsilon^{(E)}={}-\fracmm{1}{H_E^2}\fracmm{dH_E}{dt_E}
\end{equation}
and
\begin{equation}\label{eta}
\eta^{(E)}=\varepsilon^{(E)}-\fracmm{1}{2\,\varepsilon^{(E)}\,H_E}\fracmm{d\varepsilon^{(E)}}{dt_E}~.
\end{equation}

After the differentiation of Eq.~(\ref{epsilon}) as
\begin{equation}\label{depsilon}
\fracmm{d\varepsilon^{(E)}}{dt_E}=2H_E\,\left(\varepsilon^{(E)}\right)^2-\fracmm{1}{H^2_E}\fracmm{d^2H_E}{dt_E^2}~,
\end{equation}
we find
\begin{equation}
\eta^{(E)}={}-\fracmm{1}{2\,H_E}\fracmm{\fracmm{d^2H_E}{dt_E^2}}{\fracmm{dH_E}{dt_E}}\,~.
\end{equation}

Since
\begin{equation}
\label{d2HE}
\fracmm{d^2H_E}{dt_E^2}={}-\fracmm{1}{M^2_\mathrm{Pl}}\fracmm{d^2\phi}{dt_E^2}\,\fracmm{d\phi}{dt_E}\,~,
\end{equation}
we also get
\begin{equation}\label{etaHdphi}
    \eta^{(E)}={}-\fracmm{1}{H_E\fracmm{d\phi}{dt_E}}\fracmm{d^2\phi}{dt_E^2}~.
\end{equation}

Next, substituting Eq.~\eqref{etaHdphi} into Eq.~\eqref{fieldEq} yields the inflaton equation of motion in the form~\footnote{Equation (\ref{equetaphi}) in the case of the ultra-slow-roll (USR) regime with  $V_{E,\phi}\longrightarrow 0$ implies $\eta^{(E)}\longrightarrow 3$. In the literature, the normalization of the SR parameter $\eta^{(E)}$ sometimes differs by the sign and the factor of 2.}

\begin{equation}
\label{equetaphi}
H_E\fracmm{d\phi}{dt_E}\left(3-\eta^{(E)}\right)+V_{E,\phi}=0~.
\end{equation}

In the Einstein frame, the CMB observables are given by the amplitude of scalar perturbations $A_s$, the scalar spectral index $n_s$, the tensor-to-scalar ratio $r$, and their running, $\alpha_s$ and $\alpha_t$, respectively.  In the SR approximation defined by {\it small \/} SR parameters $\varepsilon^{(E)}$ and $\eta^{(E)}$, the CMB observables in the {\it leading} approximation with respect to the SR parameters evaluated at the horizon crossing are (see, e.g., Ref.~\cite{Toyama:2024ugg} for details about the relation between the CMB power spectrum and its tilts)
\begin{equation}
\label{infpaparEinstein}
\begin{split}
& n_s\approx 1-2\varepsilon^{(E)}+\fracmm{d\ln(\varepsilon^{(E)})}{dN_{E}}=1-4\varepsilon^{(E)}+2\eta^{(E)}\,,  \quad \alpha_s \approx -\fracmm{d\,n_s}{dN_{E}}\,,\\
& r\approx 16\varepsilon^{(E)}\,, \qquad  A_s\approx \fracmm{2H_E^2}{\pi^2 M_{\mathrm{Pl}}^2 r^{(E)}}\,,\quad
\alpha_t \approx \fracmm{1}{8}\fracmm{d\,r}{dN_{E}}\,,
\end{split}
\end{equation}
in terms of e-folds $dN_E=-H_Edt=-dk/k$ or scale $k$. This is
consistent with $N_{E}=-\ln(a_E/a_{E_0})$, where $a_E(t)$ is the FLRW
scale factor in the Einstein frame and the constant $a_{E_0}$ is the
value of $a_{E}$ at the end of inflation. Hence, $N_E$ is a
monotonically decreasing function during inflation and $N_E=0$
corresponds to the end of inflation. The Planck pivot scale $k_*$ of
the horizon crossing is taken at $k_*=0.05~{\rm Mpc}^{-1}$, which
corresponds to $N_{E_{*}}=56\pm 8$ for Starobinsky
inflation~\cite{Toyama:2024ugg}.

\subsection{Jordan frame}

The Hubble flow (SR)  parameters in the Jordan frame are defined by
\begin{equation}
\label{eps}
\varepsilon_1={}-\fracmm{\dot{H}}{H^2}=\fracmm{1}{2}\,\fracmm{d\ln(H^2)}{dN}\,, \qquad
\varepsilon_n=\fracmm{\dot{\varepsilon}_{n-1}}{H{\varepsilon}_{n-1}}={}-\fracmm{d\ln(\varepsilon_{n-1})}{dN}~.
\end{equation}

We find it useful to introduce the additional SR parameters in our setup, which are given by
\begin{equation}
\label{zeta}
\zeta_1=\fracmm{\dot{F}_{,R}}{HF_{,R}}=\fracmm{F_{,R R}\dot{R}}{H F_{,R}}={}-\fracmm{d\ln(F_{,R})}{dN}\,,\quad
\zeta_n=\fracmm{\dot{\zeta}_{n-1}}{H{\zeta}_{n-1}}={}-\fracmm{d\ln(\zeta_{n-1})}{dN}\,.
\end{equation}

Using Eq.~\eqref{SYSTEMFRN}, the first SR parameters can be expressed in terms of $H^2$ and $R$ as
\begin{equation}
\label{eps1HR}
\varepsilon_1=2-\fracmm{R}{6H^2},\quad \zeta_1=\fracmm{\left(6H^2-R\right)F_{,R}+F}{6H^2F_{,RR}\left(12H^2-R\right)}-\fracmm{R}{3H^2}~.
\end{equation}

Equation~(\ref{YR}) can be rewritten to the form
\begin{equation}
\label{Yzeta}
Y=\fracmm{M_{\mathrm{Pl}}H}{\sqrt{2F_{,R}\,}}\left(1+\fracmm{1}{2}\zeta_1\right)~,
\end{equation}
and Eq.~(\ref{equY}) can be rewritten to
\begin{equation}
\label{equzeta1}
3M_{\mathrm{Pl}}^4H^2\left(1+\fracmm{1}{2}\zeta_1\right)^2=AF_{,R}{\dot{R}}^2+2F_{,R}V_{\mathrm{eff}}~.
\end{equation}
Both sides of this equation do not vanish because both $V_{\mathrm{eff}}$ and $F_{,R}$ are positive for all relevant values of $R$ during inflation.

By using the relations
\begin{equation}
\label{DHeTe}
\fracmm{dH_E}{dt_E}=\fracmm{M_{\mathrm{Pl}}\dot{Y}}{\sqrt{2\,F_{,R }}}=\fracmm{M_{\mathrm{Pl}}^2 H^2}{8F_{,R }}\left(2\zeta_1\zeta_2-(2\varepsilon_1+\zeta_1)(2+\zeta_1)\right)\,,
\end{equation}
 one can connect the Hubble flow parameters in the Einstein frame to the Hubble flow  parameters in the Jordan frame as follows~\cite{Pozdeeva:2025ied}:
\begin{equation}\label{varepsilonE1}
\begin{split}
\varepsilon^{(E)}&={}-\fracmm{1}{H_E^2}\,\fracmm{d{H}_E}{d{t}_E}=\varepsilon_1+\fracmm{\zeta_1\left(1-\varepsilon_1\right)}{2+\zeta_1}-\fracmm{2\zeta_1\zeta_2}{\left(2+\zeta_1\right)^2}\\
&\approx \varepsilon_1+\fracmm{1}{2}\zeta_1-\fracmm{\zeta_1}{4}\left(\zeta_1+2\zeta_2+2\varepsilon_1\right),
\end{split}
\end{equation}
where we have kept the second-order terms in $\zeta_1$.

Using Eq.~(\ref{equ01}) in the following form:
\begin{equation}
\label{equ01param}
   \zeta_1={}-2\varepsilon_1+\zeta_1\left(\zeta_1+\zeta_2-\varepsilon_1\right)~,
\end{equation}
we obtain
\begin{equation}
\label{varepsilonE2}
\varepsilon^{(E)}\approx \varepsilon_1+\fracmm{1}{2}\zeta_1-\fracmm{\zeta_1}{4}\left(\zeta_1+2\zeta_2+2\varepsilon_1\right)=\fracmm{1}{4}\zeta_1\left(\zeta_1-4\varepsilon_1\right)\approx 3\varepsilon_1^2+{\cal O}\left(\varepsilon_1^3\right).
\end{equation}

Similarly, by using Eqs.~(\ref{eta}) and (\ref{varepsilonE2}), we get
\begin{equation}
   \eta^{(E)}=\varepsilon^{(E)}+\fracmm{1}{2}\,\fracmm{d\ln(\varepsilon^{(E)})}{dN_E}\approx{}-\varepsilon_2\fracmm{dN}{dN_E}+3\varepsilon_1^2+{\cal
O}\left(\varepsilon_1^3\right)~.
\end{equation}

The relation
\begin{equation}
\label{dNdNE}
\fracmm{dN}{dN_E}=\fracmm{2}{2+\zeta_1}\approx 1+\varepsilon_1
\end{equation}
gives
\begin{equation}
   \eta^{(E)}\approx{}-\varepsilon_2-\varepsilon_1\varepsilon_2+3\varepsilon_1^2+{\cal
O}\left(\varepsilon_1^3\right)\,.
\end{equation}

Therefore, the CMB observables in the Einstein frame can be related to the Hubble flow parameters in the Jordan frame as follows:
\begin{eqnarray}
\label{Jtilts}
r\!&\approx&\! 16\varepsilon^{(E)}\approx 4\left|\zeta_1\left(\zeta_1-4\varepsilon_1\right)\right|\approx 48\varepsilon_1^2\,, \label{rslr}  \\
  n_s\!&\approx&\! 1-4\varepsilon^{(E)}+2\eta^{(E)}  \approx  1-2\varepsilon_2-6\varepsilon_1^2-2\varepsilon_1\varepsilon_2 \approx  1-2\varepsilon_2\,,
  \label{nsslr} \\
  \alpha_s\!&\approx&\! {} - 2\varepsilon_2\varepsilon_3 ,\qquad  \alpha_t \approx  2~\fracmm{d\,\varepsilon^{(E)}}{dN_{E}}\approx
 {}- 12\varepsilon^2_1\varepsilon_2. \label{alphaslr}
\end{eqnarray}
where we have kept only the leading contributions. The amplitude of scalar perturbations is given by
\begin{equation}
\label{As}
A_s=\fracmm{H_E^2}{8\pi^2 M_{\mathrm{Pl}}^2\varepsilon^{(E)}}\approx\fracmm{H^2}{48\pi^2 F_{,R}\varepsilon^2_1}\approx  \fracmm{H^2}{\pi^2 rF_{,R}}\,.
\end{equation}

Of course, Eqs.~(\ref{Jtilts})--(\ref{As}) make sense only in the SR approximation, because the higher-order terms with
respect to $\varepsilon^{(E)}$ and $\eta^{(E)}$ are ignored.

\subsection{USR regime}

Unlike the SR regime, in the USR regime  assuming $\fracmm{A}{2}{\dot{\s}}^2\ll V_{\mathrm{eff}}$ does not imply
\begin{equation}
A\ddot\s+\fracmm{A_{,\s}}{2}{\dot{\s}}^2 \ll V_{\mathrm{eff},\s}
\end{equation}
because the field derivative of the potential $V_{\mathrm{eff}}$ almost vanishes. Instead, Eq.~(\ref{KGequVeff}) reads
\begin{equation}
\label{KGequURS}
\ddot\s={}-3Y\sqrt{\fracmm{2F_{,R}}{M_{\mathrm{Pl}}^2}}\dot{\s}-\fracmm{A_{,\s}}{2A}{\dot{\s}}^2.
\end{equation}

The USR regime can be realized by engineering a nearly-inflection point in the inflaton potential below the inflation scale. It follows from Eq.~(\ref{equ00}) that at such (de Sitter-like) point one has
\begin{equation}\label{Hds}
    H_{dS}^2=\fracmm{F(R_{dS})}{6F_{,R}(R_{dS})}\quad {\rm and,~hence,}\quad R_{dS}=\fracmm{2F(R_{dS})}{F_{,R}(R_{dS})}~.
\end{equation}
Under the condition (\ref{restr2}), Eq.~(\ref{Hds}) is equivalent to
\begin{equation}
\label{DVeffdS}
     V_{eff,\s}(\s_{dS})=0   \quad {\rm and,~hence,} \quad  F_{,\s}(\s_{dS})=\fracmm{2F(\s_{dS})}{\s_{dS}}\,.
\end{equation}

When $V_{eff,\s\s}(\s_{dS})=0$, one also has
\begin{equation}
\label{D2VeffdS}
 F_{,\s\s}(\s_{dS})=\fracmm{2F(\s_{dS})}{\s_{dS}^2}\,.
\end{equation}

An example of the USR regime at the end of the Starobinsky inflation leading to PBH production was proposed and studied  in Refs.~\cite{Saburov:2023buy,Saburov:2024und}.

By using the SR parameter $\eta = M^2_{\rm Pl}V_{E,\phi\phi}/V_E$ in terms of the inflaton potential in the Einstein frame
$V_E(\phi)=\fracmm{M^4_{\rm Pl}}{4F_{,R}^2}(F_{,R}R(\phi)-F(R(\phi)))$ , with
$F_{,R}=\fracmm{M^2_{\rm Pl}}{2}\exp\left(\sqrt{\fracmm{2}{3}}\fracmm{\phi(R)}{M_{\mathrm{Pl}}}\right)$, we find
\begin{equation}
V_E=
\fracmm{M^2_{\mathrm{Pl}}}{2}R\exp\left(-\sqrt{\fracmm{2}{3}}\fracmm{\phi}{M_{\mathrm{Pl}}}\right)-F\exp\left(-2\sqrt{\fracmm{2}{3}}\fracmm{\phi}{M_{\mathrm{Pl}}}\right)~,
\end{equation}
\begin{equation}\label{Vprime}
    V_{E,\phi}=
    \sqrt{\fracmm{2}{3}\,}{M_{\mathrm{Pl}}}^{-1}\left({F}\exp\left(-2\sqrt{\fracmm{2}{3}}\fracmm{\phi}{M_{\mathrm{Pl}}}\right)-V_E\right)~,
\end{equation}
and
\begin{equation}
V_{E,\phi\phi}=
\sqrt{\fracmm{2}{3}}\left[\fracmm{M_{\mathrm{Pl}}}{2}\exp\left(-\sqrt{\fracmm{2}{3}}\fracmm{\phi}{M_{\mathrm{Pl}}}\right)\fracmm{dR}{d\phi}  -\fracmm{2F}{M^2_{\mathrm{Pl}}}\sqrt{\fracmm{2}{3}}\exp\left(-2\sqrt{\fracmm{2}{3}}\fracmm{\phi}{M_{\mathrm{Pl}}}\right)-\fracmm{V_{E,\phi}}{M_{\mathrm{Pl}}}\right]~.
\end{equation}

When $\eta$ is close to $3$, the first derivative of the potential almost vanishes, $V_{E,\phi}\approx 0$, and the potential itself becomes
\begin{equation} \label{USRV}
V_E\approx F\exp\left(-2\,\sqrt{\fracmm{2}{3}}\fracmm{\phi}{M_{\mathrm{Pl}}}\right)\approx \fracmm{F}{F_{,R}^{2}}~.
 \end{equation}

\section{More on SR approximation in Jordan frame}

Equation (\ref{equY}) in the SR approximation is just the Friedmann equation,
\begin{equation}
\label{equYslr}
3M_{\mathrm{Pl}}^2Y^2 \approx V_{\mathrm{eff}}\,.
\end{equation}
After differentiating this equation with respect to the cosmic time in the Jordan frame and using Eq.~(\ref{Fr21Y}), we obtain
\begin{equation}
\label{phiequappr}
\dot{R}\approx{}-\fracmm{M_{\mathrm{Pl}}V_{\mathrm{eff},R}}{3YA\sqrt{2F_{,R}}}\,~.
\end{equation}
This equation can be rewritten as
\begin{equation}
\label{equchislr}
{R\,}'\approx\fracmm{M_{\mathrm{Pl}}V_{\mathrm{eff},R }}{3YHA\sqrt{2F_{,R}}}=\fracmm{M_{\mathrm{Pl}}V_{\mathrm{eff},R}Y}{3Y^2HA\sqrt{2F_{,R}}}\approx\fracmm{M_{\mathrm{Pl}}^4V_{\mathrm{eff},R }}{2F_{,R}AV_{\mathrm{eff}}}
\left(1-\fracmm{F_{,RR}}{2F_{,R}}{R\,}'\right)\,.
\end{equation}

After solving Eq.~(\ref{equchislr}) for ${R\,}'$, we get the SR evolution equation for the Ricci scalar $R$ in the Jordan frame
as
\begin{equation}\label{RNslr}
    {R\,}'\approx {} - \fracmm{2F_{,R}\left(RF_{,R}-2F\right)}{F_{,RR}\left(2RF_{,R}-F\right)}\,.
\end{equation}

In addition, we get
\begin{equation}
\label{zeta1R}
    \zeta_1(R)\approx\fracmm{2\left(RF_{,R}-2F\right)}{2RF_{,R}-F}
\end{equation}
and
\begin{equation}\label{H2F}
    H^2(R)\approx\fracmm{\left(2RF_{,R}-F\right)^{2}}{54F_R\left(RF_R-F\right)}~.
\end{equation}

Equation (\ref{eps1HR}) also implies
\begin{equation}
\label{eps1R}
\varepsilon_1=2-\fracmm{R}{6H^2}\approx 2-\fracmm{9F_RR\left(RF_R-F\right)}{\left(2RF_{,R}-F\right)^{2}}\,~.
\end{equation}

Using Eqs.~(\ref{zeta1R}) and (\ref{eps1R}), we find in the SR approximation that
\begin{equation}
\label{eps1zeta1R}
\varepsilon_1(R)={}-\fracmm{1}{2}\zeta_1(R)+\fracmm{1}{4}\zeta_1^2(R).
\end{equation}

Then, by using Eqs.~(\ref{RNslr}), (\ref{zeta1R}) and (\ref{eps1R}), we get
\begin{equation}
\label{eps2R}
\varepsilon_2(R)={}-\fracmm{1}{\varepsilon_1}\fracmm{d\varepsilon_1}{dN}={}-\fracmm{{R\,}'}{\varepsilon_1}\fracmm{d\varepsilon_1}{dR}\\
\approx\fracmm{18F F_{,R}\left(F F_{,RR} R- F_{,R}^{2}R+F_{,R}F\right) }{F_{,RR}\left( 4 F_{,R}^{3}R^{3}-3F^{2}F_{,R}R + F^{3}\right)}~,
\end{equation}
\begin{equation}
\label{zeta2R}
\zeta_2(R)\approx\fracmm{6F_{,R}\left(\left[FF_{,RR}-F_{,R}^2\right]R+F_{,R}F\right)}{F_{,RR}\left(2F_{,R}R-F\right)^2}~,
\end{equation}
and
\begin{equation}
\label{zeta3R}
\begin{split}
    \zeta_3(R)&\approx {}
    -\fracmm{4\left( F_{,R}R-2\,F  \right)}{F_{,RR}^{2}\left(F_{,RR}F R- F_{,R}^{2}R+ F_{,R}F\right) \left(2F_{,R}R-F\right)^{2}}\\
   & \times  \left( F_{,RR} \left[\fracmm{F}{2}F_{,RR}^{2}R \left(2 F_{,R} R+F\right) -\fracmm{F}{2}F_{,R}F_{,RR}\left(2 F_{,R}R-3F\right)
   \right.\right.\\
   &\left.\left. {} + F_{,R}^{3}\left(F- F_{,R}R\right)  \right] - \fracmm{1}{2}F_{,RRR} F_{,R}^{2} \left(  F_{,R}R-F  \right)  \left(2F_{,R}R-F\right)\right)\,.
\end{split}
\end{equation}

Equation (\ref{eps1zeta1R}) in the same approximation implies
\begin{equation}
\fracmm{d\epsilon_1}{dN}\approx-\fracmm{1}{2}\fracmm{d\zeta_1}{dN}+\fracmm{1}{4}\fracmm{d\zeta_1^2}{dN}\approx\fracmm{\zeta_1\zeta_2}{2}-\fracmm{\zeta_1^2\zeta_2}{2}\approx\fracmm{\zeta_1}{2}\left(\zeta_2-\zeta_1\zeta_2\right)~.
\end{equation}
Hence, we also have
\begin{equation}
\epsilon_2\approx\zeta_2(1-\zeta_1)\approx \zeta_2
\end{equation}
and
\begin{equation}
\epsilon_3\approx\zeta_3-\fracmm{\zeta_1\zeta_2}{1-\zeta_1}\approx\zeta_3.
\end{equation}

We use the simplified relations $\varepsilon_1\approx{}-\zeta_1/2$, $\varepsilon_2\approx\zeta_2$, and $\varepsilon_3\approx\zeta_3$ to calculate the inflationary parameters by Eqs.~(\ref{rslr})--(\ref{alphaslr}),
and Eq.~(\ref{eps1R}) to get the value of $R$ corresponding to the end of inflation.
Substituting those results into the right-hand-side of Eq.~(\ref{Jtilts}) gives
\begin{equation}
\label{rR}
    r\approx48\varepsilon_1^2 \approx 12\zeta_1^2 \approx 48\left(\fracmm{RF_{,R}-2F}{2RF_{,R}-F}\right)^2
\end{equation}
and
\begin{equation}
\label{nsR}
\begin{split}
    n_s\approx 1-2\zeta_2\approx1-\fracmm{12F_{,R}\left(\left[FF_{,RR}-F_{,R}^2\right]R+F_{,R}F\right)}{F_{,RR}\left(2F_{,R}R-F\right)^2}~,
\end{split}
\end{equation}
and allows us to calculate the running indices as
\begin{equation}
\label{alphasR}
\alpha_s\approx - 2\epsilon_2\epsilon_3\approx - 2\zeta_2\zeta_3
\end{equation}
and
\begin{equation}
\alpha_t\approx - 12\epsilon_1^2\epsilon_2\approx - 3\zeta_1^2\zeta_2~,
\end{equation}
 where we have kept only the leading terms. Also, we obtain
\begin{equation}
\label{AsR}
    A_s\approx  \fracmm{H^2}{\pi^2 rF_{,R}}\approx {\fracmm { \left( 2 F_{,R} R-F  \right)^4}{2592\,{\pi }^{2} F_{,R} ^{2}\left(F_{,R} R-F\right)\left(  F_{,R} R-2F \right)^2}}\,.
\end{equation}

In terms of the functions $U(\s)$ and $V(\s)$, Eqs.~(\ref{RNslr}) and (\ref{H2F}) take the form
\begin{equation}\label{chislr}
    {R\,}'(\s)\approx\fracmm{2M_{\mathrm{Pl}}^{4}UV_{\mathrm{eff},\s }}{M_{\mathrm{Pl}}^4U_{,\s}V_{\mathrm{eff},\s}+4AU^2V_{\mathrm{eff}}}={}-\fracmm{2U\left(2VU_{,\s}-V_{,\s}U \right)}{U_{,\s}\left(VU_{,\s}+V_{,\s}U\right)}
\end{equation}
and
\begin{equation}\label{H2sigma}
    H^2(\s)\approx\fracmm{\left(VU_{,\s}+V_{,\s}U\right)^{2}}{54VUU_{,\s}^{2}}\,~,
\end{equation}
which are similar to the relations obtained in~Ref.~\cite{Pozdeeva:2025ied} for the single-field models of inflation with a non-minimally coupled scalar field.

\section{Starobinsky model in Jordan frame}

It is instructive to apply the equations, obtained in the preceding sections,  to the well-known case of the Starobinsky inflation model described by Eq.~(\ref{starS}), without resorting to the Einstein frame or the corresponding inflaton potential (\ref{starV}).

Given
\begin{equation}
F(R)=\fracmm{M^2_{\rm Pl}}{2} \left( R +\fracmm{R^2}{6\,m^2}\right)~, \label{starF}
\end{equation}
we find in the SR approximation that
\begin{equation}
H^2\approx \fracmm{\left(2{m}^{2}+R \right)^{2}}{12\left(3\,{m}^{2}+R\right)} \approx \fracmm{R}{12}
\end{equation}
and
\begin{equation}
\label{dRdNStar}
{R\,}'\approx\fracmm{4{m}^{2} \left( 3{m}^{2}+R \right) }{2m^2+R} \approx 4m^2~,
\end{equation}
when $R\gg m^2$.

After solving Eq.~(\ref{dRdNStar}), we get the running e-folds as
\begin{equation}
\label{NR}
N=\fracmm{R}{4m^2}-\fracmm{1}{4}\ln \fracmm{3m^2+R}{R_0}\,,
\end{equation}
where $R_0$ is an integration constant.

The SR parameters of the Starobinsky model  in the (original) Jordan frame can be expanded during inflation
with respect to the small variable
\begin{equation}\label{SparSR}
z = \fracmm{m^2}{R}.
\end{equation}

Using $z\ll 1$, we find the leading terms as follows:
\begin{equation}
\label{zeta1Rst}
    \zeta_1(z)={}-4{\fracmm {z}{2z+1}}\approx {} - 4z~,
\end{equation}
\begin{equation}
\label{zeta2Rst}
\zeta_2(z)=4{\fracmm {z \left( 3z+1 \right) }{ \left( 2z+1 \right) ^{2}}}
\approx 4z~,
\end{equation}
\begin{equation}
\label{zeta3Rst}
\zeta_3(z)=4{\fracmm {z \left( 4\,z+1 \right) }{ \left( 2\,z+1 \right) ^{2}}}\approx 4z~,
\end{equation}
\begin{equation}
\label{eps1Rst}
    \varepsilon_1(z)=2\,{\fracmm {z \left( 4z+1 \right) }{ \left( 2z+1 \right) ^{2}}}\approx 2z~,
\end{equation}
\begin{equation}
\label{eps2Rst}
\varepsilon_2(z)=4{\fracmm {z \left( 6z+1 \right)  \left( 3z+1 \right) }{ \left( 2
z+1 \right) ^{2} \left( 4z+1 \right) }}\approx 4z~,
\end{equation}
and
\begin{equation}\label{eps3Rst}
\varepsilon_3(z)=4{\fracmm {z \left( 108{z}^{3}+74\,{z}^{2}+16z+1 \right) }{
 \left( 2z+1 \right)^{2} \left( 4z+1 \right)  \left( 6z+1 \right) }}\approx 4z\,.
\end{equation}
Hence, in the leading order, we get from Eqs.~(\ref{zeta1Rst})--(\ref{eps2Rst}) that
\begin{equation} \label{SRappr}
\zeta_1 \approx -2\varepsilon_1~, \quad \zeta_2 \approx \varepsilon_2 \approx 2\varepsilon_1, \quad \zeta_3 \approx \varepsilon_3 \approx 2\varepsilon_1~.
\end{equation}

Using Eqs.~(\ref{rR}) and (\ref{nsR}), we also obtain
\begin{equation}
n_s\approx 1 -2\zeta_2=1-8z\fracmm{\left( 3z+1 \right) }{\left( 2z+1 \right)^{2}}
\end{equation}
and
\begin{equation}
r\approx 12\zeta_1^2=\fracmm{192z^2}{\left( 2z+1 \right)^{2}}\,.
\end{equation}
Hence, the leading terms in Eqs.~(\ref{Jtilts})--(\ref{As}) are given by
\begin{equation} \label{JtiltsS}
r\approx 48 \varepsilon_1^2 +{\cal O}(\varepsilon_1^3)~,\quad n_s \approx 1- 4\varepsilon_1 +
{\cal O}(\varepsilon_1^2)~,
\end{equation}
\begin{equation}\label{alphaSR}
\alpha_s \approx {} -8 \varepsilon_1^2 +{\cal O}(\varepsilon_1^3)~,\quad \alpha_t \approx {}-24\varepsilon_1^3 +
{\cal O}(\varepsilon_1^4)~.
\end{equation}

The formal end of inflation corresponds to $R_{\rm end}$ arising from the condition $\varepsilon_1(R_{\rm end})=1$, which yields
\begin{equation}
      R_{\rm end}\approx (\sqrt{5}-1)m^2 \approx 1.2 m^2~.
\end{equation}

The initial value $R_{\rm in}$ of the spacetime scalar curvature at the beginning of inflation is related to the duration of inflation measured by e-folds $N_*=N_{\rm in}-N_{\rm end}$. Our results for the Starobinsky model  are summarized in Table~\ref{R2inflation}, while they agree with those derived in the Einstein frame, see, e.g., Ref.~\cite{Ivanov:2021chn}.

\begin{table}[h]
\begin{center}
\caption{The values of $n_s$, $r$, $R_{\rm in}$, and $N_*$ in the Starobinsky model.\label{R2inflation}}
\begin{tabular}{|c|c|c|c|c|c|}
  \hline
  $n_s$ & $0.961$ & $0.964$ & $0.969$ & $0.971$ & $0.974$\\ \hline
  $r$ & $0.0043$ & $0.0037$ & $0.0027$ & $0.0024$ & $0.0019$\\ \hline
  $R_{in}/m^2=1/z_{in} $ &  $209.6$ & $226.7$ & $262.6$ & $280.4$ & $312.2$  \\ \hline
  $N_* $ &$51$ & $55$ & $64$ & $69$ &$77$\\
  \hline
\end{tabular}
\end{center}
\end{table}

In the leading order of the SR approximation, we find from Eq.~(\ref{NR}) that
\begin{equation}
\label{efoldsS}
      N_* \approx \fracmm{1}{4z} \approx  \fracmm{1}{2\varepsilon_1}~.
\end{equation}
Then Eq.~(\ref{JtiltsS}) reproduces the well-known results for the Starobinsky model,
\begin{equation}
n_s=1-\fracmm{2}{N_*}+{\cal{O}}\left(N_{*}^{-2}\right)\,,\qquad r=\fracmm{12}{N_{*}^2}+{\cal{O}}\left(N_{*}^{-3}\right)\,,
\end{equation}
and
\begin{equation}
\alpha_s={}-\fracmm{2}{N^2_*}+{\cal{O}}\left(N_{*}^{-3}\right)\,,\qquad  \alpha_t={}-\fracmm{3}{N_{*}^3}+{\cal{O}}\left(N_{*}^{-4}\right)\,.
\end{equation}

As is clear from Table~\ref{R2inflation}, it is possible to increase the value of $n_s$ in the Starobinsky model  by increasing the duration of inflation measured by the e-folds beyond $70$ that is too high to be acceptable.
It is impossible to get a positive running $\alpha_s$ without a modification of the model.

\section{ACT versus Planck}

Let us compare the Planck/BICEP/Keck observations \cite{Planck:2018jri,BICEP:2021xfz} of the CMB radiation with more recent CMB observations due to Atacama Cosmology Telescope \cite{ACT:2025fju,ACT:2025tim} combined with
DESI 2024 data \cite{DESI:2024mwx}.

The Planck/BICEP data gives the following values for  the key CMB observables related to the power spectrum of scalar perturbations~\cite{BICEP:2021xfz} :
\begin{equation} \label{PLB}
n_s = 0.9651 \pm 0.0044\,, \quad \alpha_s = -0.0069 \pm 0.0069~,
 \end{equation}
whereas the ACT/DESI data implies~\cite{ACT:2025fju,ACT:2025tim,DESI:2024mwx}
\begin{equation}\label{ACT}
n_s = 0.9743 \pm 0.0034\,, \quad \alpha_s = 0.0062 \pm 0.0052~.
\end{equation}

The ACT/DESI data does not significantly change the upper bound on $r$ and the value of $A_s$~\cite{Galloni:2022mok},
\begin{equation}
\label{Inflparamobserv}
A_s=(2.10\pm 0.03)\times 10^{-9}\qquad {\rm and} \qquad  r < 0.028~.
\end{equation}

Thus the ACT/DESI data favours a {\it higher} scalar spectral index $n_s$ with small {\it positive} running $\alpha_s$ and a {\it concave} (not convex) inflaton potential. It is worth mentioning that the ACT/DESI data should be taken with a grain of salt because it itself has tension between the data about baryon acoustic oscillation (BAO) from DESI under the assumption of the standard cosmological model and the CMB data  \cite{Ferreira:2025lrd}, though the difference in the most cases is just about $2\sigma$.

Nevertheless, it makes sense to confront the existing single-field models of inflation with the ACT constraints given above
and ask for a possible cure. It is not difficult to increase the value of $n_s$, while keeping the desired duration of inflation within $50\div 60$ e-folds,  by adding extra terms to the known  models and adjusting extra parameters to obtain an ACT-conforming inflaton potential during inflation in the SR approximation in the Einstein frame, see e.g., Refs.~\cite{Addazi:2025qra,Gialamas:2025ofz,Kallosh:2025rni,Antoniadis:2025pfa,Aoki:2025wld,Berera:2025vsu,Dioguardi:2025vci,Salvio:2025izr,Kim:2025dyi,Gao:2025onc,Wolf:2025ecy,Pallis:2025nrv}. However, it is more difficult to change the sign of $\alpha_s$ in the SR approximation. Actually,  the majority of single-field inflation models predict negative values of $\a_s$ \cite{Martin:2024qnn} and, hence, upward bending of the inflaton potential.  The ACT-favored downward bending of the inflaton potential  for the higher values of the inflaton field before inflation is not an issue to be worried about because those values are beyond the scope of inflation.

As regards the Starobinsky inflation model in the Jordan frame, an obvious phenomenological approach to meet the ACT constraints is adding the higher order terms with respect to the spacetime scalar curvature in the context of $F(R)$ gravity. Alternatively,  from the fundamental viewpoint, string theory corrections to the gravitational effective action may do the same job~\cite{Toyama:2024ugg}. As was demonstrated in  Ref.~\cite{Addazi:2025qra},  it is possible to increase the value of $n_s$ in Starobinsky inflation to the ACT-favored region, while having small negative values of $\a_s$, by adding the $R^3$- or $R^4$- corrections to the action (\ref{starS}).

The ACT data is a more serious challenge for many single-field inflation models with PBH production because the latter tends to decrease the CMB scalar index $n_s$. Solving this problem implies the necessity for additional fine-tuning and/or more parameters~\cite{Allegrini:2024ooy,Frolovsky:2025iao}.

\section{ACT/DESI-improved model of Starobinsky inflation}

Since the ACT observations have merely $\sim 2\sigma$ tension with the predictions of the Starobinsky model, it is reasonable to study small deformations of the model in the context of $F(R)$ gravity for the limited range of the spacetime scalar curvature $R$ relevant to inflation. Since those modifications are small and conservative, they belong to the same universality class of inflationary models as the Starobinsky model itself, so that all such models can be referred to as Starobinsky inflation where the $R^2$-term is responsible for flatness of the inflaton potential during inflation.

For instance, the Starobinsky inflation models defined by
\begin{equation} \label{F3}
    F_3(R)=\fracmm{M^2_{\rm Pl}}{2}\left(R+\fracmm{R^2}{6m^2}+\fracmm{\delta_3 R^3}{36m^4}\right)
\end{equation}
and
\begin{equation}\label{F4}
    F_4(R)=\fracmm{M^2_{\rm Pl}}{2}\left(R+\fracmm{R^2}{6m^2}+\fracmm{\delta_4 R^4}{48m^6}\right)
\end{equation}
with the dimensionless parameters $\d_3$ and $\d_4$, respectively, were well studied in the literature~\cite{Berkin:1990nu,Saidov:2010wx,Kaneda:2010ut,Kaneda:2010qv,Ketov:2010qz,Huang:2013hsb,Miranda:2017juz,Ivanov:2021chn}. Though it is possible to get the values of $n_s$ in agreement with the ACT results~\cite{Addazi:2025qra,Gialamas:2025ofz},  the values of $\alpha_s$ remain negative in those $F(R)$ gravity models of inflation.

Let us consider a new model by adding an $R^5$-term as well, with
\begin{equation}\label{F5}
    F_5(R)=\fracmm{M^2_{\rm Pl}}{2}\left(R+\fracmm{1}{6m^2}R^2+\fracmm{c_3}{m^4} R^3+\fracmm{c_4}{m^6} R^4+\fracmm{c_5}{m^8}R^5\right)~,
\end{equation}
where $c_3$, $c_4$ and $c_5$ are the dimensionless coupling constants.

In this model, we find
\begin{equation}\label{zeta1_R5}
 \zeta_1= {}-\fracmm {4\left({z}^{4}-c_3{z}^{2}-2c_4z-3c_5\right)}{2{z}^{4}+{z}^{3}+10c_3{z}^{2}+14c_4z+18c_5}~.
\end{equation}
The observational bound $r<0.028$ and Eq.~(\ref{rR}) also imply
\begin{equation}\label{zeta1in}
    |\zeta_{1_{in}}|<0.048~.
\end{equation}
Hence, any term except the $R^2$ in Eq.~(\ref{F5}) should not dominate, all the parameters $c_i$ should be small, $\abs{c_i}\ll 1$, whereas a small positive value of $c_5$ can be responsible for changing the sign of $\zeta_1$.

Using the SR approximation formulae (\ref{rR}), (\ref{nsR}), and (\ref{alphasR}), the values  of $c_i$  can be chosen to meet the ACT favored values of $n_s$ and $\alpha_s$. Our results are summarized in Table~\ref{FR2345}. Those results are robust against small changes in the values of $c_i$, while $R_{in}$ can be adjusted to keep $N_{*}$ within the desired range.

The stability conditions (\ref{restr2}) are satisfied for all positive $R$. The effective potential $V_{eff}$ has a maximum at $R> 150\,m^2$, so that $R(t)$ is a monotonically decreasing function during inflation. The values of $n_s$, $r$, and $\alpha_s$ are independent upon the value of $m$ that is fixed by the observed value of~$A_s$.

\begin{table}[h]
\begin{center}
\caption{The values of $R_{\rm in}$ (in units of $m^2$),  $N_*$ and $r$ for $n_s=0.974$ and $\alpha_s=0.0062$.
\label{FR2345}}
\begin{tabular}{|c|c|c|c|c|c|}
  \hline
   $c_3$& $c_4$& $c_5$ & ${R_{\rm in}}$ & $N_{*}$ & $r$  \\
   \hline
  $3.874\times 10^{-4}$  & $-2.583\times 10^{-6}$ & $6.084\times 10^{-9}$ & $130$ &$64.2$& $0.0030$ \\
  $3.412\times 10^{-4}$  & $-2.282\times 10^{-6}$ & $5.355\times 10^{-9}$ & $130$ &$56.1$& $0.0040$ \\
  $3.073\times 10^{-4}$  & $ -2.068\times 10^{-6}$ & $4.845\times 10^{-9}$ & $130$ & $ 51.1$& $ 0.0050$ \\
  $2.805\times 10^{-4}$  & $ -1.903\times 10^{-6}$ & $4.459\times 10^{-9}$ & $130$ & $ 47.6$& $ 0.0060$ \\
  $2.377\times 10^{-4}$  & $ -1.529\times 10^{-6}$ & $3.361\times 10^{-9}$ & $140$ & $ 49.0$& $ 0.0060$ \\
 \hline
\end{tabular}
\end{center}
\end{table}

\section{Generalisations of the model for smaller scales}

The curvature corrections to the Starobinsky model (\ref{starS}), given by Eq.~(\ref{F5})  with the coefficients from Table 2, do not affect physics at other scales smaller than the scale of inflation, where those corrections are negligible. In particular, the transition to reheating and the radiation-dominated era in our model (\ref{F5}) are the same as those in the original Starobinsky model (\ref{starS}).

Cosmological inflation in the early Universe has many similarities with dark energy (DE) in the current Universe because both describe the Universe expansion with acceleration. It is possible to (phenomenologically) unify both descriptions in $F(R)$ gravity alone, though finding a viable $F(R)$-function obeying all physical and observational constraints is non-trivial. A solution to this problem was proposed by Appleby, Battye and Starobinsky (ABS) in Ref.~\cite{Appleby:2009uf}, without destroying inflation and the standard Big Bang cosmology. The $F(R)$-gravity function proposed in Ref.~\cite{Appleby:2009uf} reads
\begin{equation} \lb{Ff}
	\fracmm{2}{M^2_{\rm Pl}}F(R)=  (1+g\tanh b) R + gE_{AB}  \ln \left[\fracmm{\cosh \left(\frac{R}{E_{AB}}-b\right)}{\cosh (b)}\right] +\fracmm{R^2}{6 m^2} ~~,
\end{equation}
where $m$ is the Starobinsky mass, the ABS parameter $E_{AB}$ is defined by
\begin{equation} \lb{ABp}
	E_{AB}=\fracmm{R_0}{2g\ln(1+e^{2b})}~,
\end{equation}
while the other parameters $g$ and $b$ of the order one define the shape of the corresponding scalar potential. The new physical scale in the ABS model is given by the parameter $R_0$, whose value in the case of the present DE is determined by the current Hubble parameter, $R_0\sim H^2_0$.

Just adding our corrections beyond the $R^2$-term from Eq.~(\ref{F5}) to the ABS-proposed $F(R)$-function (\ref{Ff}) defines the unified $F(R)$-gravity model of inflation (or primordial DE) and the current DE, this time in agreement with ACT also, leading to a single scalar field in the Einstein frame. The corresponding scalar potential is only locally defined and cannot be computed analytically, though numerical calculations are possible. In contrast, in our approach in the Jordan frame, all our
$F(R)$-gravity functions in use are elementary, well-defined and smooth, while there is no necessity to go to the Einstein frame in order to get physical predictions. The $F(R)$-function of the unified model reads
\begin{equation}
	\fracmm{2}{M^2_{\rm Pl}}F(R)   =(1+g\tanh b) R + gE_{AB}  \ln \left[\fracmm{\cosh \left(\frac{R}{E_{AB}}-b\right)}{\cosh (b)}\right] +\fracmm{R^2}{6 m^2}\nonumber \end{equation}
\begin{equation} +\fracmm{c_3}{m^4} R^3+\fracmm{c_4}{m^6} R^4+\fracmm{c_5}{m^8}R^5~.\lb{uniF}
\end{equation}

Moreover, the same $F$-function (\ref{uniF}) but with the high scale $R_0\sim m^2$ under the scale of inflation can
be used to describe efficient PBHs production after Starobinsky inflation, as was demonstrated in Refs.~\cite{Saburov:2023buy,Saburov:2024und} after fine-tuning the parameters as $g\approx 2.25$ and $b\approx 2.89$.
The corresponding scalar potential has two plateaus with a nearly-inflection point between them, which implies the existence of an USR regime. The produced PBHs have masses in the asteroid-size range between $10^{16}$ g and $10^{20}$ g,
beyond the Hawking evaporation limit, so that they can be considered as viable candidates for the current dark matter. The PBHs-induced gravitational waves have the frequencies about $10^{-2}$ Hz, which makes them detectable by the future space-based gravitational interferometers such as LISA \cite{Smith:2019wny} and DECIGO \cite{Kudoh:2005as}, while
the quantum (loop) corrections to the power spectrum are small, see Refs.~\cite{Saburov:2023buy,Saburov:2024und} for a derivation of those results.

It is also possible to add the ABS terms to our model  (\ref{F5}) {\it twice} with proper values of the parameters but the different scales corresponding to  the PBHs scale and the DE scale, respectively,  in order to simultaneously describe inflation, PBHs production and current DE.  This shows the power of our minimal approach based on $F(R)$ gravity in the Jordan frame without adding extra matter. Coupling matter scalars to $F(R)$ gravity for describing ACT-consistent inflation with PBHs production was studied in  Refs.~\cite{Kim:2025dyi,Pozdeeva:2025wsl}.

Finally, as was noticed in Ref.~\cite{Appleby:2009uf}, the number of e-folds $N$ for inflation often increases in the unified
$F(R)$ gravity models, while in the case of Starobinsky inflation $N$ may increase up to $70$. In turn, it leads to an
ACT-favoured  increase of the scalar tilt $n_s$. Though it is not enough to get the desired ACT range of $n_s$, it implies less
fine-tuning of the coefficients $c_i$ in Eq.~(\ref{uniF}) that makes our model more robust.

\section{Conclusion}

Single-field models of inflation, with or without PBH production, as well as inflation models based on $F(R)$ gravity, are under scrutiny due to the CMB observations with increasing precision. On the theoretical side, the single-field models of inflation are also constrained by quantum gravity conjectures known as the Swampland program \cite{Palti:2019pca}, though those constraints do not rule out the Starobinsky inflation~\cite{Scalisi:2018eaz,Ketov:2025nkr}.

The $\mu$-type distortion \cite{Zeldovich:1969ff,Chluba:2011hw} of the CMB leads to additional constraints on the primordial power spectrum at smaller scales up to~$k\sim 10^4\,\mathrm{Mpc}^{-1}$. The $\mu$-distortion was estimated in Refs.~\cite{Chluba:2015bqa, Schoneberg:2020nyg, Unal:2020mts, Pajer:2012vz, Tagliazucchi:2023dai}.  The upper bound from COBE/FIRAS observations \cite{Fixsen:1996nj} is given by  $\mu < 9 \cdot 10^{-5}$ ($95\%\, \text{C.L.}$). As regards
small deformations of the Starobinsky model, it was found in Ref.~\cite{Frolovsky:2025qre} that   $\mu \simeq 10^{-10}$ that is significantly below the observational limit.

The recent ACT/DESI observational data disfavors the Starobinsky inflation model (\ref{starS}) by $2\sigma$ and thus motivates a search for its modifications. The simplest modifications including only one additional term proportional to $R^3$ or $R^4$ can provide the desired increased value of the cosmological tilt $n_s$ of scalar perturbations but the running index  $\alpha_s$ remains negative~\cite{Addazi:2025qra,Gialamas:2025ofz}. To obtain a positive value of $\alpha_s$, one should consider further modifications of the Starobinsky model. Generically, it leads to a complicated inflaton potential $V_E$ that can be studied only numerically. To overcome this difficulty,  our new version of the SR approximation in the Jordan frame allows one to investigate the $F(R)$ gravity models of inflation without resorting to the Einstein frame. Using this SR approximation, we have proposed and studied a new inflation model defined by Eq.~(\ref{F5}) that agrees with the ACT constraints to the CMB observables for some values of the model parameters. To the best of our knowledge, the proposed model is the first $F(R)$-gravity-based model of inflation with a positive running index $\alpha_s$ and $1\sigma$-agreement with the ACT/DESI data. The other inflationary parameters are in agreement with the observations as well.

As was demonstrated in section~9, our approach is extendable to unified $F(R)$ gravity models of inflation, PBHs production and DE, without resorting to the Einstein frame,  see Refs.~\cite{Amendola:2006we,Ribeiro:2023yhh,Oikonomou:2025qub} for comparison.

Since the difference between the Planck and ACT observations of the CMB is just about $2\sigma$, it is still premature  to dispose the Planck data. Therefore,  future precision measurements of the cosmological tilts and their running by experiments such as LiteBIRD \cite{LiteBIRD:2022cnt} and the Simons Observatory \cite{SimonsObservatory:2025wwn} are needed to resolve the current tensions between the available data.

 \section*{Acknowledgements}

S.V.K. was partially supported by Tokyo Metropolitan University and the World Premier International Research Center Initiative, MEXT, Japan. Research of E.O.P. and S.Yu.V. was conducted for the scientific program of the National Center for Physics and Mathematics, Section 5, Particle Physics and Cosmology, 2023--2025.

\bibliography{KPV_FR}{}
\bibliographystyle{JHEP}
\end{document}